\documentclass[acmsmall,screen,nonacm]{acmart}

\usepackage{booktabs}
\usepackage{tabularx}
\usepackage{array}
\usepackage{graphicx}
\usepackage{wrapfig}
\usepackage{mdframed}
\usepackage{multirow}

\newcolumntype{P}[1]{>{\raggedright\arraybackslash}p{#1}}
\newcolumntype{Y}{>{\raggedright\arraybackslash}X}

\newcommand{\lead}[1]{%
  \par\noindent\textbf{#1}\enspace
}

\newcounter{finding}
\newcommand{\finding}[1]{\refstepcounter{finding}
    \vspace{0.5mm}
    \begin{mdframed}[linecolor=gray,roundcorner=12pt,backgroundcolor=gray!15,linewidth=3pt,innerleftmargin=2pt, leftmargin=0cm,rightmargin=0cm,topline=false,bottomline=false,rightline=false]
    \textbf{Finding \arabic{finding}:} #1
    \end{mdframed}
    \vspace{0.5mm}
}

\setcopyright{none}
\renewcommand\footnotetextcopyrightpermission[1]{}

\title[When Messages Go Wrong]{When Messages Go Wrong: Understanding Communication Faults in LLM-Based Multi-Agent Systems}

\author{Jinhao Zhou}
\email{jinhao@suou.waseda.jp}
\affiliation{%
  \institution{Waseda University}
  \country{Japan}
}

\author{Xiaoyu Zhang}
\authornote{Corresponding author.}
\email{joshingrain@gmail.com}
\affiliation{%
  \institution{The Hong Kong University of Science and Technology}
  \country{Hong Kong SAR, China}
}

\author{Jiongchi Yu}
\email{jiongchiyu@acm.org}
\affiliation{%
  \institution{Nanyang Technological University}
  \country{Singapore}
}

\author{Fahao Chen}
\email{chenfh@ieee.org}
\affiliation{%
  \institution{Shandong University}
  \country{China}
}

\author{Jun Wu}
\authornotemark[1]
\email{junwuhn@sjtu.edu.cn}
\affiliation{%
  \institution{Shanghai Jiao Tong University}
  \country{China}
}

\renewcommand{\shortauthors}{Zhou et al.}

\begin{document}

\begin{abstract}
Large language model-based multi-agent systems (LLM-MAS) increasingly employ specialized agents that exchange messages to accomplish complex tasks, including repository-level code generation. As these systems move from controlled demonstrations toward real-world applications, their reliance on networked protocols, remote tools, and model endpoints exposes agent communication to message loss, delay, and context expiration. Unlike traditional distributed systems that mainly exchange structured messages, LLM-MAS frequently rely on open-ended natural language to convey instructions and evidence. Communication anomalies can therefore alter task-relevant information while leaving messages processable, allowing semantic errors such as constraint loss and plan drift to propagate through subsequent agent interactions. Although previous research has examined adversarial threats and failures arising within individual agents, systematic investigations of inter-agent communication faults remain limited. This paper presents a large-scale empirical study of how communication faults manifest, propagate, and affect task outcomes in LLM-MAS. We construct a cross-layer taxonomy linking lower-stack disturbances to application-visible faults and develop a fault-injection and tracing pipeline to examine their downstream consequences. Our evaluation covers more than 29,000 executions across 100 tasks, five domains, three workflow topologies, and three model configurations. The results reveal that 20.7\% of failures with complete detection records occur without an immediate failure signal, reaching 76.2\% for partial-content faults. Workflow topology also substantially influences fault tolerance. Flat maintains 53.0\% task success under communication faults, compared with 3.3\% and 3.2\% for Sequential and Hierarchical, respectively. This advantage primarily arises when an alternative information path preserves the evidence required by downstream agents. We further investigate execution-side and evidence protection mechanisms and find that they recover different failure cases. Evidence recovery, however, depends on whether trustworthy information remains accessible through an unaffected path. These observations offer empirical insights into communication-fault propagation, workflow resilience, and protection design for LLM-MAS. The code and experimental artifacts are available at \url{https://github.com/SonGoKu21/mas_communication/}, with an interactive companion website at \url{https://songoku21.github.io/mas_communication/}.
\end{abstract}


\maketitle

\section{Introduction}
\label{sec:introduction}

Large language models (LLMs) increasingly support multi-agent software systems in which specialized agents collaborate through message exchange~\cite{wu2023autogen,hong2024metagpt,qian2024chatdev,guo2024multiagents,he2025llmmasse}.
These LLM-based multi-agent systems (LLM-MAS) already serve tasks such as web navigation, software engineering, and office automation~\cite{zhou2024webarena,jimenez2024swebench,xu2024agentcompany}.
As LLM-MAS move toward real-world deployment, reliable agent communication becomes critical because communication faults can disrupt coordination, alter intermediate information, and ultimately affect task outcomes.
Understanding how these faults manifest and affect downstream execution is therefore essential for building dependable LLM-MAS.

\begin{figure}[tbp]
\centering
\includegraphics[width=1\textwidth]{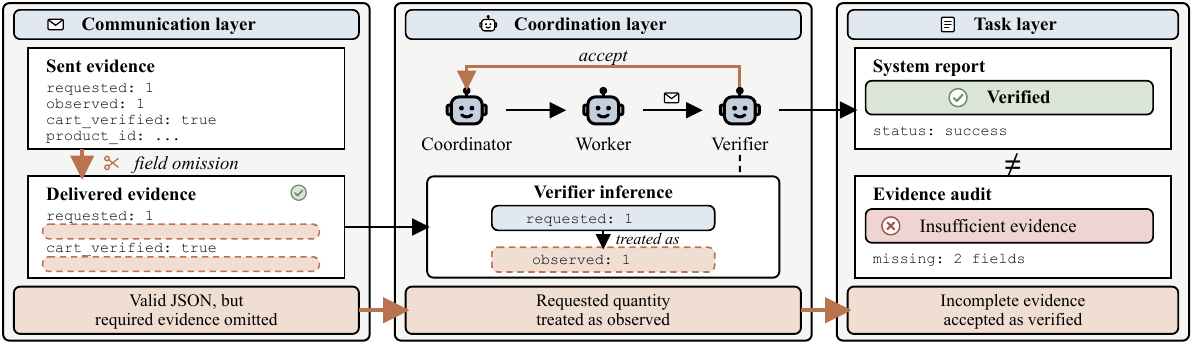}
\caption{A silent partial-content communication failure in an LLM-MAS.
Although the message remains syntactically valid, the Verifier accepts incomplete evidence without an immediate failure signal.}
\label{fig:motivating-example}
\end{figure}

Crucially, a real-world LLM-MAS rarely runs inside a single process.
Its agents coordinate over networked protocols such as the Model Context Protocol and Agent-to-Agent messaging, and they access remote tools, sandboxes, and model endpoints through these channels~\cite{wu2023autogen,anthropic2024mcp,google2025a2a}.
Emerging deployments push this further into physically distributed settings, for example multi-robot teams driven by LLM agents.
As a result, an LLM-MAS can encounter communication disturbances across different layers, including network delay and loss, connection or request timeouts, logical-message non-delivery, and context expiration~\cite{rfc9293,grpcdeadlines,grpcretry}.
Traditional distributed systems also face these issues, but structured protocols provide explicit mechanisms for detecting and handling many communication errors.
LLM-MAS, however, often exchange open-ended natural language that each agent interprets autonomously.
A communication fault can leave a message processable while altering the information consumed downstream.
This is particularly important in LLM-MAS, where later agents depend on requests, observations, and evidence produced earlier in the workflow.
As illustrated in \autoref{fig:motivating-example}, a partial delivery can remove required evidence fields while leaving the message syntactically valid, allowing the workflow to continue without an immediate failure signal.
Existing research has mainly examined adversarial threats, internal agent failures, coordination errors, and broader reliability issues~\cite{wang2026dataagents,he2025red,2025arXiv251204129L,cemri2025mast,zhu2025multiagentbench,zhang2025failureattribution,tan2026agentchaos}.
Systematic studies of faults in inter-agent communication remain limited.

To the best of our knowledge, we conduct the first empirical study dedicated to communication faults in LLM-MAS.
We formalize a cross-layer communication-fault taxonomy that connects established lower-stack communication anomalies, such as packet loss and stream interruption, to application-visible faults such as message unavailability and incomplete delivery.
Building on this taxonomy, we design a fault-injection and tracing pipeline that perturbs inter-agent communication during MAS execution and records how the resulting effects manifest in downstream workflow traces.
Our study covers more than 29,000 executions across 100 tasks, five domains,
three workflow topologies, and three model configurations.
Among the scheduled fault executions, 15,882 reach their target communication
boundaries and are successfully applied.
We additionally conduct targeted source-mapping and position-comparison
experiments, together with a 1,098-execution main protection study and
follow-up analyses of recovery conditions.
Our investigation is driven by four overarching Research Questions (RQs):

\(\bullet\)
\noindent
\textbf{RQ1 (Manifestation \& Propagation): How do communication faults become visible to agents and propagate to MAS consequences and final task outcomes?}
Across 3,780 source-mapping executions, lower-stack disturbances exhibit a many-to-many mapping to agent-visible faults.
Among 15,882 applied faults, 89.4\% are associated with recorded downstream MAS consequences and 80.2\% end in task failure.
Meanwhile, 1,457 executions still succeed after a recorded consequence.
Among 8,335 failures with complete detection records, 20.7\% remain silent.
These results separate fault manifestation, downstream consequence, outcome, and failure visibility.

\(\bullet\)
\noindent
\textbf{RQ2 (Sensitivity): How do injection position and task context shape the impact of communication faults?}
Matched comparisons show substantial differences across communication stages.
In Admin, Flat retains success rates from 94.4\% to 100.0\% under I4 non-delivery across the three models.
Sequential and Hierarchical both fall to 0\% at I4, although they retain 34.7\% to 75.9\% success at I3.
The impact of message loss therefore depends on the information carried at each stage and whether subsequent execution can obtain the missing information.

\(\bullet\)
\noindent
\textbf{RQ3 (Fault Tolerance): How do topologies and model configurations affect outcomes under communication faults?}
Matched comparisons show that Flat gains most of its advantage at I4.
Compared with Sequential and Hierarchical, Flat achieves absolute success-rate gains of 66.5-66.6\% under I4 non-delivery and 41.3-42.3\% under I4 corruption, while differences at I2 and I3 remain below 0.7\%.
Across 1,408 common faulted units, changing to Flat recovers 19.3\% to 27.0\% of baseline failures, whereas changing the LLM recovers only 0.2\% to 4.5\%.
The two changes recover largely different failure cases, showing that workflow structure and model configuration provide different forms of fault tolerance.

\(\bullet\)
\noindent
\textbf{RQ4 (Protection Mechanisms): Which mechanisms protect execution and information dependencies against communication faults?}
We conduct a 1,098 execution study comparing execution-side and evidence protection, while targeted follow-up analyses examine recovery conditions and overhead.
The two mechanisms recover complementary failure sets.
Evidence recovery remains effective while the workflow retains access to an unaffected information path, but fails when the same fault reaches every available evidence path.

This paper makes the following contributions:

\begin{itemize}

\item \textbf{A cross-layer communication-fault taxonomy for LLM-MAS.}
We organize lower-stack communication faults, agent-visible message faults, and downstream MAS consequences within a cross-layer taxonomy.
The taxonomy distinguishes fault origin from what a receiving agent observes and further captures faults introduced by agent-runtime handling.

\item \textbf{A fault-injection and tracing pipeline for inter-agent communication.}
We build a pipeline that injects faults at concrete communication boundaries, verifies their application-visible manifestation, and follows each affected communication dependency through downstream workflow state and task outcome.

\item \textbf{A large-scale empirical characterization of communication faults.}
Across 100 tasks, five domains, three topologies, and three deployed model configurations, we quantify fault manifestation, sensitivity, and tolerance under controlled communication faults.
Targeted follow-up studies further characterize silent failures, protection mechanisms, and the conditions required for successful recovery.

\item \textbf{An open-source release.}
We publicly release the necessary experimental results, scripts, artifacts, and an interactive companion website to support reproduction and future research on communication reliability in LLM-MAS.\footnote{\url{ https://songoku21.github.io/mas_communication/}}

\end{itemize}

\section{Background \& Related Work}
\label{sec:background}

\subsection{LLM-MAS}
\label{sec:mas-systems}

\begin{figure}[tbp]
\centering
\includegraphics[width=0.92\textwidth]{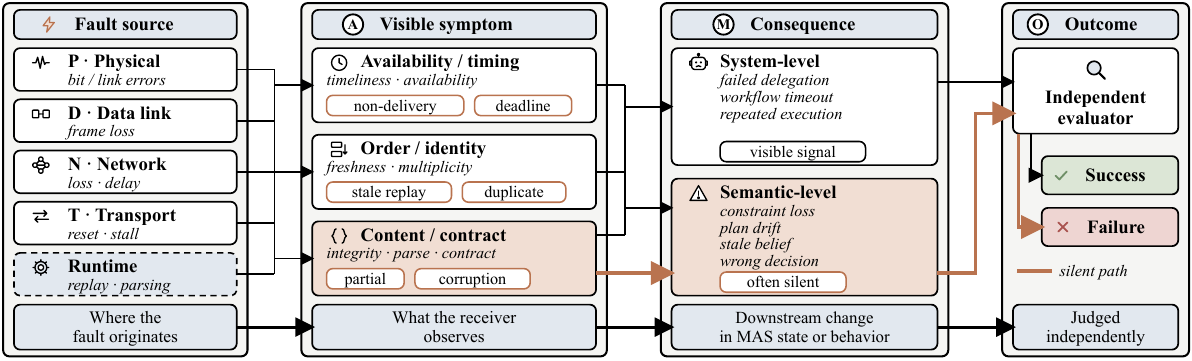}
\caption{Cross-layer communication-fault propagation. P/D/N/T denote conventional lower-stack communication layers; A represents application-level inter-agent communication and receiver-visible communication faults; M and O represent downstream MAS consequences and task outcomes.}
\label{fig:layers}
\end{figure}

An LLM-based multi-agent system (LLM-MAS) consists of specialized, model-backed agents that collaborate through message exchange to solve complex tasks~\cite{wu2023autogen,hong2024metagpt,qian2024chatdev,guo2024multiagents,he2025llmmasse}.
Although existing frameworks adopt different orchestration mechanisms, for example, AutoGen supports message-oriented interaction~\cite{wu2023autogen}, while MetaGPT emphasizes role-based collaboration~\cite{hong2024metagpt,qian2024chatdev}, they share a common communication dependency: a request, observation, tool result, or piece of evidence produced by one agent may become the input to a later workflow step.
Such communication may further span remote tools, model endpoints, and physically distributed environments through protocols such as the Model Context Protocol (MCP) and Agent-to-Agent (A2A) messaging~\cite{anthropic2024mcp,google2025a2a}.
Following standard communication reference models~\cite{iso7498,rfc1122},~\autoref{fig:layers} illustrates this communication path and its downstream effects.
P/D/N/T denote the conventional physical, data-link, network, and transport layers, respectively; A denotes application-level inter-agent communication, while M and O denote downstream MAS consequences and final task outcomes.
We characterize communication-fault propagation through four analytical stages: the \textit{source} where a disturbance originates, the receiver-visible \textit{symptom} exposed at A, the downstream MAS \textit{consequence} at M, and the final task \textit{outcome} at O.
These stages describe the propagation of a communication fault through an LLM-MAS rather than additional protocol layers.

Formally, we model an LLM-MAS as
\begin{equation}
\Sigma=\langle G,\mathcal{B},\mathcal{T},\mathcal{S},
\mathcal{C},\mathcal{R},\mathcal{E}\rangle,
\end{equation}
where $G$ is the communication graph, $\mathcal{B}$ the deployed agents,
$\mathcal{T}$ the tools, $\mathcal{S}$ the workflow states,
$\mathcal{C}$ the receiver contracts, $\mathcal{R}$ the message-handling policies,
and $\mathcal{E}$ the task environment.
We represent a logical communication \textit{carrier} as
\begin{equation}
m_t=\langle id,src,dst,task,session,round,payload,schema,version,time\rangle.
\end{equation}
Application-layer operations such as dispatching, queueing, association, ordering, and structured-message processing determine how a carrier reaches and is consumed by a receiving agent.
Recent empirical studies mainly characterize failures arising from agent reasoning, system design, verification, or individual agent components~\cite{cemri2025mast,zhu2025multiagentbench,zhang2025failureattribution}.
In contrast, we treat communication faults along the message path as first-class causes and study how their effects propagate from receiver-visible symptoms to downstream MAS behavior and task outcomes.

\subsection{Communication Faults}
\label{sec:communication-faults}

Traditional networked and distributed systems have long treated communication faults as a major reliability concern, including loss, delay, reordering, duplication, corruption, and connection interruption~\cite{iso7498,rfc1122}. Communication stacks use checksums, retransmission, timers, duplicate handling, and redundancy to preserve reliable execution under such faults~\cite{stone2000checksum,rfc9293,10.1145/2080.357392,avizienis2004dependability}. Similar problems also arise in service-oriented systems, where timeouts and partial failures can propagate across remote-call dependencies~\cite{zhou2021microservice,heorhiadi2016gremlin,basiri2016chaos}.

LLM-MAS introduce an additional challenge because successful delivery does not guarantee that the receiving agent obtains valid task information. A message can remain parseable while carrying incomplete, stale, misassociated, or corrupted content, allowing the affected information to propagate through later workflow steps. Recent studies have reported related communication and information faults in LLM-based agent systems~\cite{tan2026agentchaos,jia2026mas}. Following the end-to-end principle~\cite{saltzer1984end}, application-level validation is therefore important when task semantics determine whether delivered information remains valid. These characteristics motivate our cross-layer taxonomy and the fault-injection analysis in \autoref{sec:methodology}.

\subsection{Fault Injection}
\label{sec:fault-injection-background}

Many communication faults are too rare, timing-dependent, or unsafe to reproduce by waiting on a live deployment, so fault injection deliberately introduces them at selected interfaces and observes the downstream effects~\cite{arlat1990fault}.
This makes the space of failure conditions controllable and repeatable, which is why fault injection has become a standard way to evaluate the resilience of dependable systems.
Early techniques emulate hardware and I/O faults in processor and memory state without special equipment~\cite{carreira1998xception,arlat1990fault}, and later system-level tools carry the same idea into distributed software.
FATE injects failures at system and library I/O points to explore combinations of faults an application may face~\cite{gunawi2011fate}, lineage-driven fault injection targets the dependencies whose disruption is most likely to invalidate an outcome~\cite{alvaro2015ldfi}, and network emulators such as NetEm reproduce loss, delay, duplication, and corruption on real traffic~\cite{hemminger2005netem}.
At the service level, Gremlin perturbs inter-service messages to test microservice resilience~\cite{heorhiadi2016gremlin}, and chaos engineering injects controlled faults into production systems to check that they tolerate realistic failures~\cite{basiri2016chaos,zhou2021microservice}.

For agent systems, AgentChaos injects faults by modifying LLM API responses~\cite{tan2026agentchaos}, while MAS-FIRE includes application-level communication faults in a broader LLM-MAS fault taxonomy~\cite{jia2026mas}.
Seyedghorban et al.~\cite{seyedghorban2026observability} combine OpenTelemetry-based tracing with runtime fault injection across agent, tool, and LLM boundaries, and ReliabilityBench~\cite{gupta2026reliabilitybench} evaluates tool-using agents under controlled API and execution failures.
Our study focuses specifically on communication faults within LLM-MAS.
We use a cross-layer taxonomy and matched executions to analyze their manifestations, stage sensitivity, fault tolerance, silent failures, and protection mechanisms across multiple task domains.
\autoref{sec:methodology} formalizes this distinction through our taxonomy and fault-injection pipeline.

\section{Methodology}
\label{sec:methodology}

Our methodology consists of two components. The taxonomy consolidates communication-fault phenomena from prior work and organizes them along the cross-layer path introduced in \autoref{sec:mas-systems}. Based on this taxonomy, the MAS fault-injection pipeline instantiates selected receiver-visible faults at controlled communication boundaries and traces how their effects propagate through downstream MAS state and task outcomes. The pipeline consists of two phases: fault generation and propagation evaluation.

\subsection{Taxonomy}
\label{sec:taxonomy}
We construct the taxonomy from 31 communication standards, protocol
specifications, and prior studies on dependability and LLM-based agents
~\cite{avizienis2004dependability,iso7498,rfc1122,anthropic2024mcp,
google2025a2a,cemri2025mast,tan2026agentchaos,jia2026mas}.
We extract 57 fault and effect descriptions and consolidate them into 43
candidate concepts.
The three researchers include one with at least five years of experience in
communication systems and two with at least two years of experience in
LLM-based agents and multi-agent systems.
After a 10-candidate pilot, two researchers independently code all candidates
by source layer, receiver-visible manifestation, and downstream MAS consequence.
Cohen's $\kappa$ ranges from 0.79 to 0.86 across the three dimensions before
a third researcher resolves disagreements, yielding 29 final categories.
We retain mappings supported by source evidence and validate selected
reproducible lower-stack mappings in our testbed.
Conditional mappings are preserved when the same source fault produces
different effects under different runtime conditions.
As shown in \autoref{tab:fault-taxonomy}, P/D/N/T denote lower-stack sources,
A receiver-visible faults, and M downstream MAS consequences.

The distinction between these groups separates fault origin from
receiver-visible manifestation and downstream consequence.
P/D/N/T characterize where a communication disturbance originates, whereas A
characterizes what the receiving agent actually observes.
An A-layer fault may result from a lower-stack disturbance or originate directly
in agent-runtime communication, for example during message delivery, ordering,
replay, association, or structured-message handling.
The source-to-A relationship is therefore conditional.
A lower-stack disturbance may produce no tracked receiver-visible manifestation,
while different sources may expose the same receiver-visible communication
fault.
M differs from both P/D/N/T and A because it records a trace-linked change in
MAS state or behavior after affected information is accepted or used.
We distinguish system-level consequences, such as failed delegation, workflow
timeout, or repeated execution, from semantic-level consequences, such as
constraint loss, incomplete aggregation, plan drift, stale-belief acceptance,
or incorrect collective decisions.
Errors originating purely from local model reasoning, parameters, activations,
or training state remain outside the scope of this taxonomy.

\begin{table}[tbp]
\caption{Cross-layer taxonomy and representative sources. P/D/N/T entries are lower-stack communication-source families; A entries are receiver-visible communication-fault families; and M entries are downstream MAS consequence groups. Representative sources for the underlying phenomena are listed in the final column.}
\label{tab:fault-taxonomy}
\footnotesize\setlength{\tabcolsep}{2.5pt}
\renewcommand{\arraystretch}{0.96}
\begin{tabularx}{\textwidth}{@{}P{.05\textwidth}P{.275\textwidth}Y P{.17\textwidth}@{}}
\toprule
Layer & Fault / effect name & Brief description & Representative source(s) \\
\midrule
P & P-F1 Bit-level corruption & Isolated or burst errors in communicated bits. & \cite{stone2000checksum} \\
  & P-F2 Link/endpoint interruption & NIC, cable, or power disruption removes a path. & \cite{rfc1122,linuxbonding} \\
  & P-F3 Physical-medium interference & Interference degrades reception and delivery timing. & \cite{rfc3819} \\
  & P-F4 Timing instability & Clock drift affects communication timing decisions. & \cite{rfc5905} \\
  & P-F5 Communication-state corruption & Hardware errors affect communication buffers or runtime state. & \cite{carreira1998xception,linuxedac} \\
\midrule
D & D-F1 Frame loss/local drop & Frames are not delivered across a local link. & \cite{rfc3366,rfc3819} \\
  & D-F2 Frame corruption/integrity failure & Damaged frames fail integrity checks or escape detection. & \cite{rfc3366,stone2000checksum} \\
  & D-F3 Duplicate/reordered delivery & Frame order or multiplicity changes. & \cite{hemminger2005netem} \\
  & D-F4 MTU/fragmentation failure & Size handling or reassembly disrupts delivery. & \cite{rfc2923} \\
  & D-F5 Contention/link instability & Contention or flapping delays or interrupts service. & \cite{rfc3819} \\
\midrule
N & N-F1 Packet loss & Independent or burst loss affects a packet path. & \cite{rfc1122,hemminger2005netem} \\
  & N-F2 Path performance degradation & Delay, jitter, bandwidth limits, or congestion slow delivery. & \cite{rfc5681,hemminger2005netem} \\
  & N-F3 Ordering/multiplicity anomaly & Packets arrive out of order or more than once. & \cite{hemminger2005netem} \\
  & N-F4 Reachability/routing failure & Partitioning, asymmetry, or filtering blocks a path. & \cite{rfc1122} \\
  & N-F5 Endpoint resolution failure & Name resolution prevents endpoint discovery. & \cite{rfc1035} \\
\midrule
T & T-F1 Connection/session interruption & Reset or interruption breaks an active exchange. & \cite{rfc9293} \\
  & T-F2 Deadline/expiration failure & A delivery or validity budget expires. & \cite{rfc5482,grpcdeadlines} \\
  & T-F3 Retry/retransmission anomaly & Repeated attempts amplify load or logical work. & \cite{rfc9293,grpcretry} \\
  & T-F4 Stream completeness failure & An exchange ends before complete content arrives. & \cite{rfc9293,rfc9112} \\
  & T-F5 Transport pressure/stall & Backlog or flow-control pressure prevents progress. & \cite{rfc5681,rfc9293} \\
\midrule
A & Timeliness & A logical message arrives late or misses its deadline. & \cite{grpcdeadlines,heorhiadi2016gremlin} \\
  & Availability & A required logical message is not delivered. & \cite{heorhiadi2016gremlin,zhou2021microservice} \\
  & Semantic integrity & A processable message carries changed or misassociated meaning. & \cite{tan2026agentchaos,he2025red} \\
  & Parseability/completeness & A message cannot be parsed or lacks required content/evidence. & \cite{anthropic2024mcp} \\
  & Multiplicity/idempotency & A logical request or delivery is repeated. & \cite{grpcretry,heorhiadi2016gremlin} \\
  & Ordering/freshness & The receiver observes displaced, out-of-order, or stale state. & \cite{rfc9293,tan2026agentchaos} \\
  & Contract validity & Structured keys, types, schema, or relations violate the expected contract. & \cite{anthropic2024mcp,google2025a2a} \\
\midrule
M & System-level consequences & Execution or coordination degrades, e.g., failed delegation or workflow timeout. & \cite{cemri2025mast,zhu2025multiagentbench,zhang2025failureattribution} \\
  & Semantic-level consequences & Accepted evidence, plans, beliefs, or decisions degrade. & \cite{cemri2025mast,he2025red,2025arXiv251204129L} \\
\bottomrule
\end{tabularx}
\end{table}

\subsection{MAS Fault Injection Pipeline}
\label{sec:pipeline}

\lead{Phase I: Fault generation.}
For each task, workflow, model configuration, and repetition, the pipeline selects a fault operator, its parameters, and a communication boundary for injection.
We define four communication boundaries, I1, I2, I3, and I4.
I1 denotes the initial task dispatch from the Controller to the workflow entry agent and is excluded because disrupting this boundary can prevent the collaborative workflow from starting.
The pipeline therefore injects faults at three boundaries during workflow execution.
\textit{Action requests} (I2) carry instructions or tool requests required for subsequent work.
\textit{Observations} (I3) return environment or tool results to the workflow.
\textit{Evidence handoffs} (I4) transfer generated results or supporting evidence to a downstream verifier or decision agent.
The selected boundary, communication edge, carrier identity, and task/session association jointly determine the communication dependency affected by the intervention.

An interceptor operates after sender-side carrier production and before receiver-side consumption. For operator $f$ with parameters $\theta$, it transforms the communication carrier, delivery behavior, or relevant A-layer runtime state as
\begin{equation}
\phi_{f,\theta}(m_t,S_t)\rightarrow(m'_t,S'_t),
\end{equation}
where $m_t$ is the original communication carrier and $S_t$ denotes the relevant runtime/workflow state. Delivery interception implements omission, delay, or duplication; queue and replay handling implement ordering/freshness faults; message association implements semantic-integrity faults; and structured key/type processing implements contract-validity faults. The intervention is restricted to the communication path and does not modify local model reasoning or tool semantics; model-provider requests and responses also remain outside the injector. Clean executions traverse the same instrumentation path using an identity operation, so the experimental configuration is held fixed apart from the selected communication intervention.

\lead{Phase II: Propagation evaluation.}
The evaluation phase first verifies whether the scheduled intervention reaches
the receiving application boundary by comparing the original and delivered
carrier or the corresponding delivery-event trace.
If execution terminates before the targeted communication event, we mark the
intervention as not applied.
This separates scheduled faults from realized exposures and ensures that
downstream analysis includes only faults that reach the intended boundary.
For each applied fault, the pipeline separately evaluates application exposure
($A$), downstream MAS consequence ($M$), final task outcome ($O$), and the
auxiliary handling-response flag ($B$).
$A$ is determined by the observed message or delivery change, independent of
final task success.
The trace then follows the affected communication dependency through accepted
state, actions, aggregation, verification, and termination.
We assign an $M$ consequence only when the trace shows a downstream state or
behavioral change associated with the affected dependency.
For semantic consequences, the altered information must also be accepted or
used by a downstream agent.
For missing carriers, the trace identifies the action or state that cannot
complete because the required communication is unavailable.
The task evaluator determines $O$ independently of $A$ and $M$.
$B$ records a handling response after fault exposure.
We retain it as an auxiliary observation because it does not establish complete
detection, correction, and state restoration.
Each execution records the communication event, original and delivered carrier
projections, message and session metadata, timing and order, accepted state,
actions, termination, and evaluator outcome.
These records support the subsequent analyses of manifestation, propagation,
sensitivity, fault tolerance, and protection mechanisms.

\section{Experimental Evaluation}
\label{sec:experiments}

\subsection{Experimental Setup}
\label{sec:experimental-design}

\lead{Tasks and datasets.}
We evaluate communication faults on 100 tasks drawn from five task domains:
WebArena Shopping, WebArena Admin, WebArena Reddit~\cite{zhou2024webarena}, SWE-bench~\cite{jimenez2024swebench}, and
TheAgentCompany (TAC)~\cite{xu2024agentcompany}.
These workloads cover interactive web tasks, administrative operations,
social-platform reasoning, repository-level software engineering, and
multi-step workplace tasks, providing diverse communication and evidence
dependencies.
Each task is executed under matched clean and faulted conditions so that
fault-induced effects can be separated from failures already present in the
clean execution.
Task-specific adapters provide the final success endpoint; for evidence-based
workflows, this endpoint may additionally require that the final result is
supported by the expected delivered evidence rather than only matching the
native benchmark answer.

\begin{figure}[t]
\centering
\includegraphics[width=\textwidth]{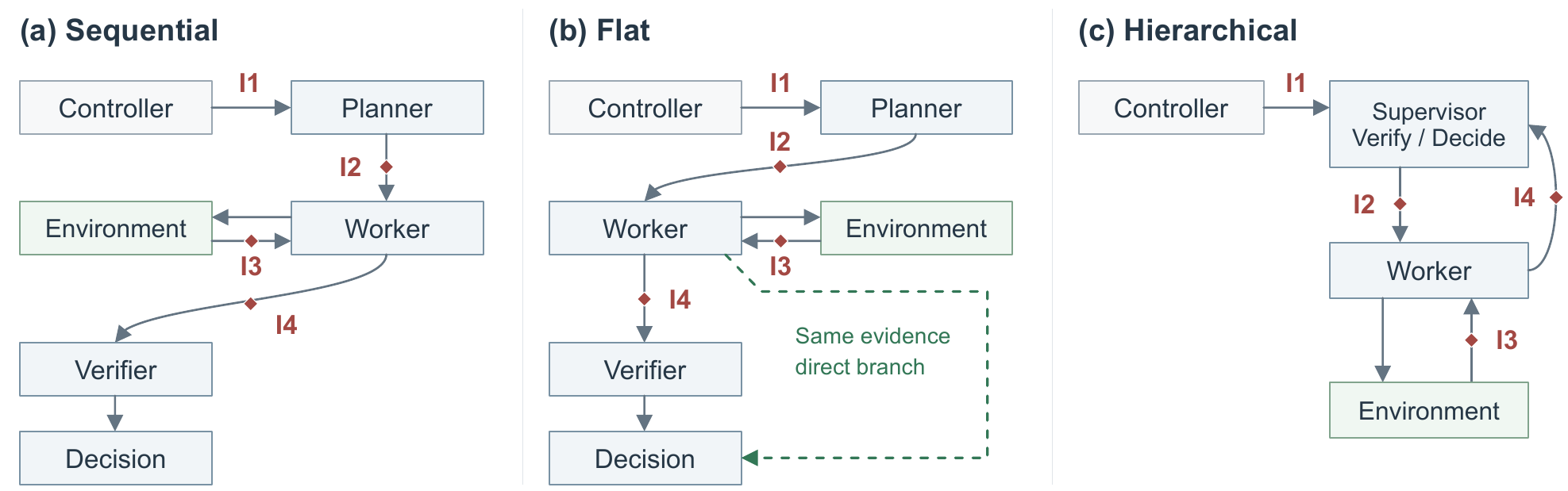}
\Description{Sequential uses one evidence path, Flat adds an alternate evidence
route, and Hierarchical aggregates evidence through a supervisor. The injection boundaries are action requests, observations, and evidence handoffs.}
\caption{AutoGen workflows and injection boundaries used in the
confirmation matrix.}
\label{fig:topology-comparison}
\end{figure}

\lead{MAS configurations.}
We instantiate three AutoGen workflow structures, shown in
\autoref{fig:topology-comparison}.
\emph{Sequential} follows a single communication path from Planner to Worker,
then through a Verifier to the final Decision.
\emph{Flat} retains this path but adds an alternate direct branch that forwards
the same pre-interception worker evidence to the Decision.
\emph{Hierarchical} replaces the separate Planner-Verifier-Decision chain with
a Supervisor that delegates the task to the Worker and subsequently receives
the worker evidence for verification and decision.
The workflows use three deployed model configurations:
DeepSeek V4 Flash~\cite{xu2026deepseek}, Qwen3.8-27B~\cite{qwen38}, and
Qwen3.5-9B~\cite{qwen35}.
Topology comparisons match executions on task, model configuration, fault
condition, injection position, and repetition, while model-configuration
comparisons additionally hold the topology fixed.
Because model provider and serving environment vary with the deployed
configuration, we interpret these settings as configuration-level rather than
isolated model-size comparisons.

\lead{Faults and injection positions.}
We inject receiver-visible communication faults at I2-I4, as defined in
\autoref{sec:pipeline} and illustrated in \autoref{fig:topology-comparison}.
I1 is excluded because disrupting the initial task dispatch can prevent
inter-agent execution from starting. The primary confirmation matrix contains one clean condition, I2 non-delivery,
and five I4 faults: valid-partial delivery, semantic corruption, stale replay,
deadline timeout, and non-delivery.
Across 100 tasks, three models, three topologies, and three repetitions, this
yields $18{,}900$ executions.
Among 16,200 scheduled faulted executions, 15,882 reach the target boundary and
receive the fault; the remaining 318 are excluded from fault-effect analyses. We also construct a position-comparison matrix containing clean execution,
non-delivery at I2-I4, and semantic corruption at I3-I4.
It contains 16,200 scheduled cells, including 10,800 shared with the primary
matrix and 5,400 additional I3 cells.
RQ2 uses the non-delivery subset, while \autoref{fig:rq3-topology}(b) uses all
five fault conditions for the topology analysis.
One scheduled cell ends in retry exhaustion, so this count denotes scheduled
cells rather than completed executions.

\lead{Metrics.}
We evaluate fault behavior at the application, MAS, and task levels.
\emph{A exposure} indicates that the scheduled intervention produces the
intended receiver-visible message or delivery change.
\emph{M propagation} indicates a trace-linked downstream change in MAS state or
behavior after the affected dependency is processed.
\emph{O outcome} records the final task-level success or failure independently
of A and M.
We additionally report the historical handling-response flag $B$ as an
auxiliary observation, but do not interpret it as a strict recovery measure.
To quantify degradation relative to executions that would otherwise succeed,
we use clean-conditional loss:
\begin{equation}
L_{\mathrm{clean}}=
\frac{\#\{\text{applied fault fails; matched clean succeeds}\}}
{\#\{\text{applied fault has a successful matched clean}\}}.
\end{equation}
We define a \emph{silent-failure rate} for applied faults that continue downstream without immediate rejection or termination and ultimately fail. The rate is reported in RQ1.
A detection record is complete when all fields required to determine fault
handling and failure visibility are present in the execution trace.
\lead{Execution protocol.}
Clean and faulted executions use the same instrumented workflow, with an identity operation as the clean control.
Each combination of task, configuration, and condition is executed three times.
Reported rates pool all included executions across the three repetitions, while per-repetition rates and their sample standard deviations assess execution variability.
Fault-effect analyses include only executions in which the scheduled fault is applied.
Additional execution matrices, inclusion criteria, and sample sizes are reported with the corresponding RQs.

\lead{Statistical analysis.}
For the confirmation matrix, we estimate 95\% confidence intervals using 10,000 task-clustered bootstrap resamples.
Paired comparisons use task-level differences and two-sided sign-flip permutation tests.
We compute paired effect sizes and Holm-adjusted \(p\)-values within each dataset and metric family.
Detailed effect sizes, confidence intervals, and raw and adjusted \(p\)-values are provided in the research artifact.
Topology and model comparisons use matched execution cells.
Cross-domain comparisons remain descriptive because tasks from different datasets are not paired.

\subsection{RQ1. Manifestation and Propagation } 

We investigate how communication faults become visible across system layers and how they affect MAS execution.
We first analyze 3,780 source-mapping executions, including 3,600 faulted runs and 180 clean runs, to study how lower-stack fault sources appear at the application layer.
We then examine 15,882 applied application-layer faults to analyze their MAS consequences and final task outcomes.
Finally, we study failure visibility using executions with complete detection records.

\begin{figure*}[t]
    \centering
    \includegraphics[width=\textwidth]{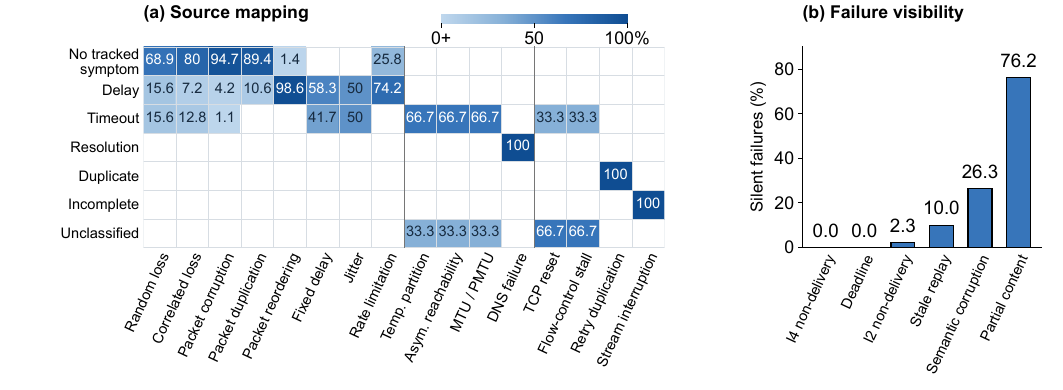}
    \caption{
    Cross-layer manifestation and failure visibility under communication faults.
    The left panel shows the distribution of primary application-level manifestations across 3,600 faulted source-mapping executions.
    The right panel shows silent-failure rates among 8,335 failures with complete detection records.
    Lower-stack fault sources exhibit many-to-many mappings to application-level symptoms, while silent failures concentrate in partial-content and semantic-corruption faults.
    }
    \label{fig:rq1-overview}
\end{figure*}

\lead{Cross-layer manifestation and downstream outcomes.}
As shown in \autoref{fig:rq1-overview} (a), lower-stack fault sources do not map one-to-one to application-level manifestations.
Across 3,600 faulted source-mapping executions, 36.0\% show no recorded consumption symptom within the tracked categories.
The remaining executions reveal substantial variation across layers.
For example, fixed delay appears as delayed delivery in 58.3\% of executions and as timeout in 41.7\%.
Jitter is evenly divided between these two manifestations.
The reverse pattern also occurs.
Timeout arises from multiple lower-stack sources, including loss, jitter, temporary partition, MTU/PMTU faults, TCP reset, and flow-control stall.
The same application-level manifestation can therefore originate from different communication conditions. The mapping also differs between faults that appear similar across layers.
Source-level packet corruption produces no recorded accepted-corruption manifestation in our mapping runs and mostly results in no tracked consumption symptom.
Likewise, lower-stack packet duplication does not directly appear as logical duplication, while retry-induced duplication consistently produces a duplicate manifestation.
These results support the separation between fault sources and agent-visible faults in our taxonomy.
The source layer identifies where a communication fault originates, while the application layer captures what the receiving agent observes.

We further examine whether an observed application-level fault determines the final task result.
Across the canonical fault categories, task failure rates range from 73.4\% for message timeout to 85.5\% for partial content.
Across all 15,882 applied faults, 89.4\% are associated with a recorded MAS consequence and 80.2\% end in task failure.
However, 1,457 executions still succeed after a recorded MAS consequence, accounting for 10.3\% of consequence-labeled runs.
Thus, an agent-visible fault or downstream MAS consequence does not uniquely determine the final outcome.

\finding{\textbf{Lower-stack fault sources do not map one-to-one to agent-visible faults or final outcomes.}
The same application-level manifestation can arise from multiple lower-stack sources, while individual sources such as delay and jitter can produce different manifestations.
Across the observed application-level fault categories, task failure rates range from 73.4\% to 85.5\%, and 1,457 executions still succeed after a recorded MAS consequence.}

\lead{Failure visibility.}
We next examine whether task failures become visible when the communication fault occurs.
Among 8,335 failures with complete detection records, 1,722 remain silent, corresponding to 20.7\%.
These executions continue without an immediate failure signal and later produce an unsuccessful task outcome.
The same 1,722 cases account for 15.6\% of the 11,022 injected runs with complete detection  records.
We exclude runs without detection records from this analysis instead of assigning them a detection outcome. 

As illustrated in \autoref{fig:rq1-overview} (b), the silent-failure rate varies substantially across fault types.
Partial-content faults show the highest rate.
Among detection-complete failures, 76.2\% of partial-content faults remain silent, compared with 26.3\% of semantic-corruption faults.
Partial-content faults alone account for 69.6\% of all silent failures in this subset.
Stale replay reaches 10.0\%, while I2 non-delivery reaches only 2.3\%.
We observe no silent failures for I4 non-delivery or deadline faults in the detection-complete subset. These differences reveal a clear pattern.
Delivery faults often remove or delay an expected message and expose an execution-level symptom.
Semantic faults can preserve a processable message while changing the information carried by that message.
The runtime can therefore continue normal message processing even though downstream agents receive incomplete or incorrect evidence.
This observation follows the end-to-end principle that successful communication at lower layers does not establish application-level correctness~\cite{saltzer1984end}. \begin{wrapfigure}{r}{0.66\textwidth}
  \centering
  \includegraphics[width=\linewidth]{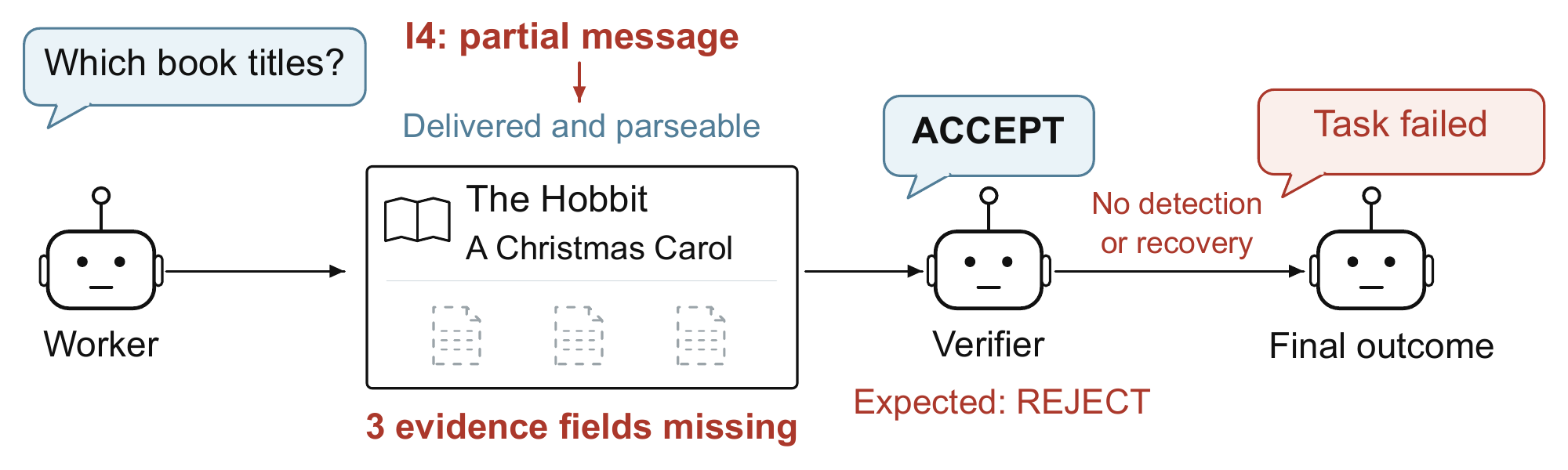}
  \caption{
Representative silent failure under partial-content delivery. The message remains parseable after required evidence is removed, and the Verifier accepts it without recorded detection or recovery.}
  \label{fig:rq1-silent-replay}
\end{wrapfigure}As shown in \autoref{fig:rq1-silent-replay}, a representative Reddit execution illustrates this pattern.
An I4 partial-content fault delivers a syntactically valid message from the Evidence Worker to the Verifier.
The message remains parseable, but three required evidence fields are missing.
The Verifier accepts the message even though the evaluation contract expects rejection.
The workflow records no detection or recovery signal and later fails.
The answer string still matches the expected answer, showing that the failure arises from the evidence contract rather than the answer text alone.

\finding{\textbf{Silent failures concentrate in faults that preserve a processable message.}
Among failures with complete detection records, 20.7\% remain silent.
The rate reaches 76.2\% for partial-content faults and 26.3\% for semantic corruption, compared with 2.3\% for I2 non-delivery and 0\% for I4 non-delivery and deadline faults.
Partial-content faults alone contribute 69.6\% of the observed silent failures.}

\subsection{RQ2. Sensitivity}
\label{sec:rq2}

RQ2 investigates why the same communication fault can produce different task outcomes.
We study two factors.
First, we compare matched non-delivery faults at I2, I3, and I4 to examine how the affected communication stage changes fault impact.
I2, I3, and I4 represent the request, state-return, and evidence-handoff stages, respectively.
Second, we compare semantic faults across five task contexts to examine how the information required by each task changes fault sensitivity.
For the stage comparison, we retain matched task-repeat units with a successful clean execution, applied faults at all three positions, and known outcomes.
For the task-context comparison, we fix the Flat topology and compare semantic corruption with partial-content faults.

\begin{wrapfigure}{R}{0.66\textwidth}
    \centering
    \vspace{-0.6em}
    \includegraphics[width=\linewidth]{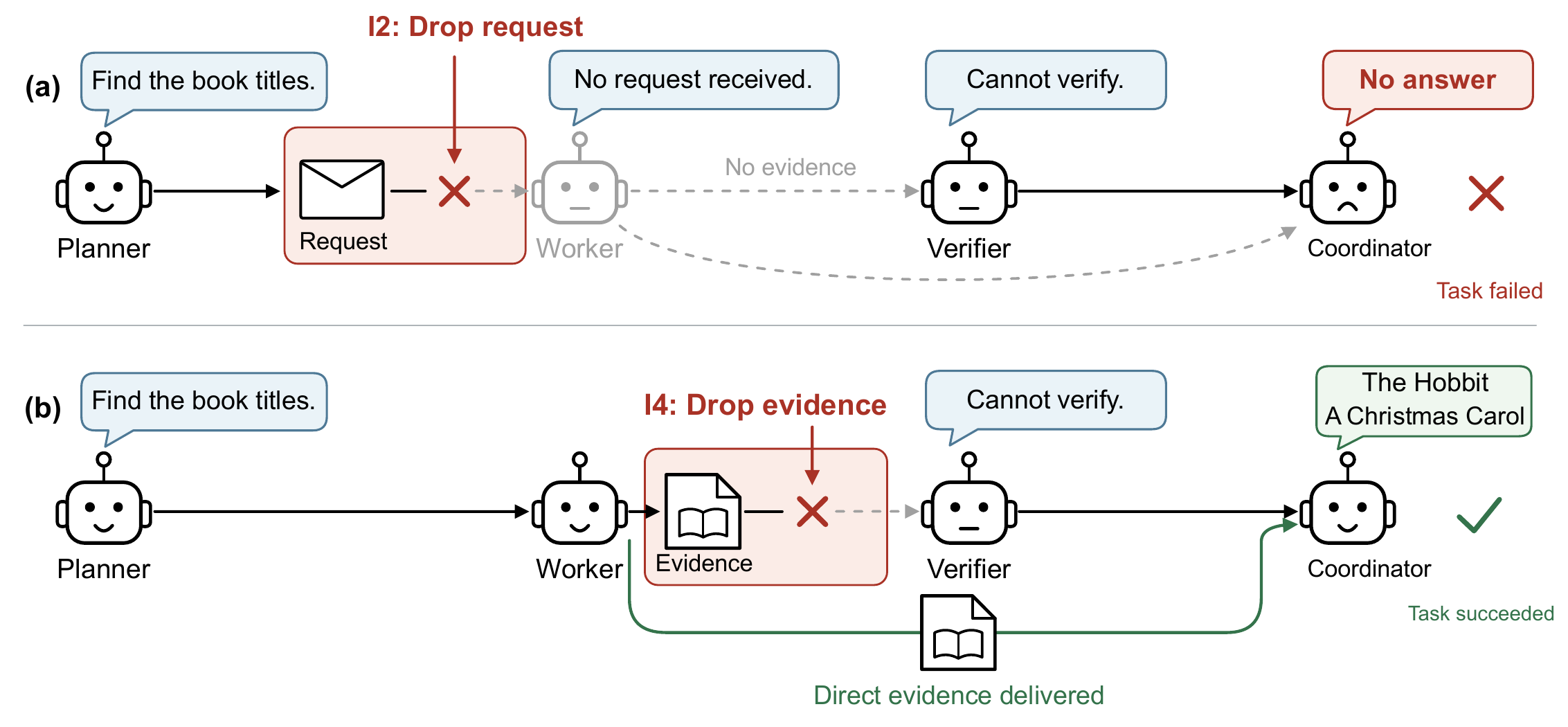}
    \caption{
    Representative impact of non-delivery at different communication stages.
    At I2, request loss prevents the Worker from producing the required evidence.
    At I4, the evidence already exists and remains available through the Flat topology's additional path.
    }
    \label{fig:rq2-position-case}
    \vspace{-0.8em}
\end{wrapfigure}

\lead{Sensitivity to communication stage.}
The matched results show no fixed severity order across I2, I3, and I4.
\autoref{tab:condition-profile} presents the clearest example in Admin.
Under Flat, I4 non-delivery retains success rates from 94.4\% to 100.0\% across the three models.
Sequential and Hierarchical fall to 0\% at the same stage.
The pattern differs at I3, where these two topologies still retain success rates from 34.7\% to 75.9\%.
The same non-delivery fault can therefore produce very different outcomes at different communication stages.
Trace analysis explains this difference.
In Admin, I3 occurs inside an iterative tool loop.
Among 110 matched DeepSeek executions under Sequential and Hierarchical, 82 receive a later table carrier after the I3 fault.
Among these 82 executions, 74 eventually succeed.
All 28 executions without a later table carrier fail.
I4 occurs after the navigation loop has finished.
None of the 110 matched I4 executions contains a later tool call.
The workflow therefore retains an opportunity to obtain another observation after I3, while this opportunity is no longer available after I4.
\begin{wraptable}{r}{0.62\columnwidth}
    \centering
    \caption{Matched task-success rates under non-delivery at I2, I3, and I4 in Admin.
    Each model uses the same matched task-repeat cohort across the three communication stages.}
    \label{tab:condition-profile}

    \small
    \setlength{\tabcolsep}{3.2pt}
    \begin{tabular}{@{}llccc@{}}
        \toprule
        Model & Topology & I2 (\%) & I3 (\%) & I4 (\%) \\
        \midrule

        \multirow{3}{*}{DeepSeek}
        & Sequential   & 90.7 & 72.2 & 0.0 \\
        & Hierarchical & 94.6 & 62.5 & 0.0 \\
        & Flat         & 87.0 & 75.9 & 94.4 \\
        \addlinespace[2pt]

        \multirow{3}{*}{Qwen3.5-9B}
        & Sequential   & 72.4 & 75.9 & 0.0 \\
        & Hierarchical & 75.0 & 75.0 & 0.0 \\
        & Flat         & 75.0 & 78.6 & 96.4 \\
        \addlinespace[2pt]

        \multirow{3}{*}{Qwen3.8-27B}
        & Sequential   & 56.2 & 35.4 & 0.0 \\
        & Hierarchical & 55.1 & 34.7 & 0.0 \\
        & Flat         & 55.1 & 34.7 & 100.0 \\

        \bottomrule
    \end{tabular}
\end{wraptable}
\autoref{fig:rq2-position-case} illustrates the same mechanism at the message level.
The task asks the agents to identify two book titles.
The paired executions use the same task, fault operator, topology, model configuration, and repetition.
When the I2 request is lost, the Worker receives no task request and produces no corresponding evidence.
The downstream agents therefore have no equivalent information to use.
When the I4 handoff is lost, the Worker has already produced the evidence.
The affected handoff removes one delivery path, while the direct Flat branch still carries the existing evidence to the Coordinator.
The final outcome therefore depends on whether the affected information can still be obtained after the communication failure. 

The other evaluated task contexts support this interpretation.
In Reddit, SWE-bench, and TAC, I2 and I3 non-delivery produce no successful matched executions under the evaluated configurations.
Our trace audit finds no later evidence reacquisition step after these faults.
At I4, the required evidence has already been generated.
Flat then retains substantial success through its additional evidence path.
Across the three models, its I4 success ranges from 92.7\% to 100.0\% on Reddit, from 78.9\% to 100.0\% on SWE-bench, and from 42.9\% to 100.0\% on TAC.
Shopping shows a weaker I4 effect, with Flat success ranging from 0\% to 53.3\%. These results show that communication stage affects fault impact through the information available afterward.
A later observation can restore information lost inside an iterative workflow.
An additional path can preserve information that already exists before a downstream handoff fails.
When neither source remains available, the missing information cannot reach subsequent agents.
This behavior is consistent with prior work showing that execution failures depend on the data dependencies affected by a fault~\cite{alvaro2015ldfi}.

\finding{\textbf{Fault impact depends on whether the affected information remains recoverable after the communication failure.}
The same non-delivery fault produces different outcomes across communication stages.
Later observations recover some I3 losses in Admin, while an additional evidence path preserves many I4 executions under Flat.}

\begin{wrapfigure}{r}{0.50\textwidth}
    \centering
    \includegraphics[width=\linewidth]{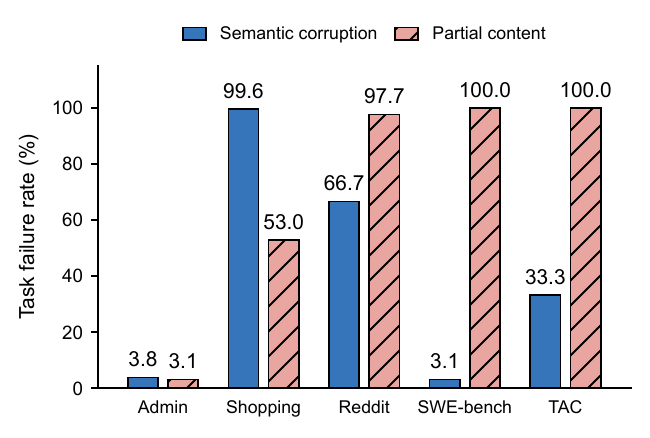}
    \caption{
    Task-failure rates under semantic corruption and partial-content faults
    across five task contexts with the Flat topology.
    }
    \label{fig:rq2-context}
\end{wrapfigure}

\lead{Sensitivity to task context.}
The same semantic fault also shows substantially different effects across task contexts.
We fix the Flat topology and compare semantic corruption with partial-content faults in \autoref{fig:rq2-context}.
Their relative severity changes sharply across the five evaluated settings. Shopping is most sensitive to semantic corruption.
Its failure rate reaches 99.6\%, compared with 53.0\% under partial content.
SWE-bench shows the opposite pattern.
Semantic corruption causes only 3.1\% failure, while partial content reaches 100.0\%.
TAC shows a similar contrast, with rates of 33.3\% and 100.0\%, respectively.
Reddit is also more sensitive to partial content, with 97.7\% failure compared with 66.7\% under corruption.
Admin remains less sensitive to both faults in this comparison. The differences follow how each evaluated task uses the affected information.
The Reddit, SWE-bench, and TAC workflows require explicit evidence for the final decision.
Partial-content faults can remove information required by this evidence contract.
Shopping follows a different execution path.
An environment action may already succeed before a later communication fault changes the evidence used for final acceptance.
Admin retains an interactive tool loop and can obtain additional observations before the final evidence handoff.
These differences change which communication faults become critical for task completion. Semantic corruption and partial-content faults exhibit different severity patterns across task contexts.
Semantic corruption dominates partial content in Shopping, while partial content dominates corruption in SWE-bench and TAC.
The impact of a semantic fault therefore depends on the information required by the task and how the workflow evaluates that information.

\finding{\textbf{The most harmful communication fault changes with task context.}
Under Flat, semantic corruption causes 99.6\% failure in Shopping but only 3.1\% in SWE-bench.
Partial-content faults show the opposite pattern, reaching 53.0\% in Shopping and 100.0\% in SWE-bench.
Fault sensitivity therefore depends on how the task uses and evaluates the affected information.}

\subsection{RQ3. Fault Tolerance}
\label{sec:rq3}

RQ3 investigates how topology and model configuration affect fault tolerance.
We first compare Flat with Sequential and Hierarchical while fixing the task context, task, repetition, fault condition, and model.
The paired analysis records both failure-to-success and success-to-failure transitions for the same execution units.
We then use 1,408 common faulted units with known outcomes across all three models and three topologies to compare topology changes with model changes.

\begin{figure*}[t]
    \centering
    \includegraphics[width=0.92\textwidth]{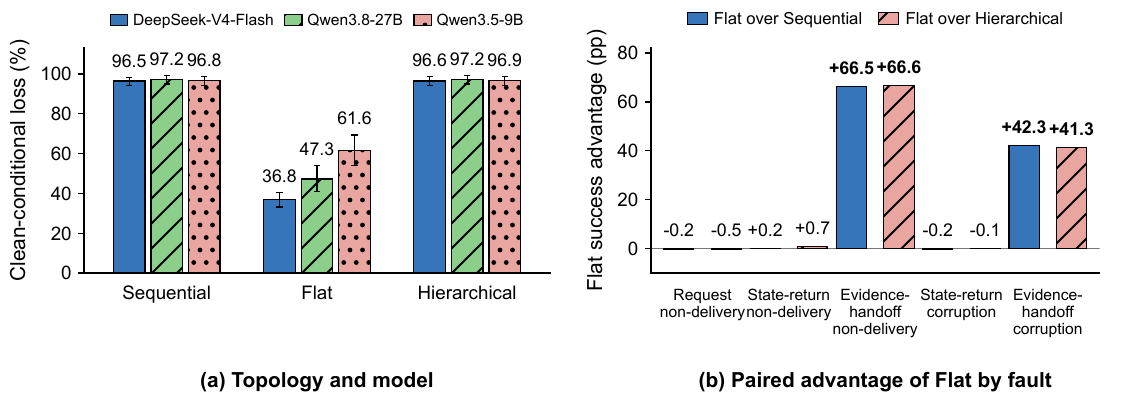}
    \caption{
    Impacts of topology and model configuration under communication faults.
(a) Clean-conditional loss across three topologies and three model configurations.
Error bars report task-clustered 95\% confidence intervals.
(b) Paired success advantage of Flat over Sequential and Hierarchical under five fault conditions.
Flat gains concentrate at the evidence handoff stage, where its branch carries the required evidence.
    }
    \label{fig:rq3-topology}
\end{figure*}

\lead{Impact of topology.}
\autoref{fig:rq3-topology}(a) shows a consistent topology pattern across the three model configurations.
Across all applied faulted executions with known outcomes, Flat retains 53.0\% task success, more than 16$\times$ the 3.3\% and 3.2\% achieved by Sequential and Hierarchical.
Consistently, Sequential and Hierarchical exhibit clean-conditional losses of 96.5-97.2\%, while Flat reduces the loss to 36.8-61.6\%.
The paired analysis in \autoref{fig:rq3-topology}(b) further shows that Flat gains most of its advantage at the evidence handoff stage.
 Under non-delivery, changing Sequential to Flat recovers 584 failed executions
without introducing new failures, corresponding to an absolute increase of
66.5\% in success rate; the increase over Hierarchical is 66.6\%.
Evidence-handoff corruption shows a similar pattern, with absolute increases
of 42.3\% and 41.3\% in success rate over Sequential and Hierarchical, respectively.
The topology difference becomes small when the fault occurs before the additional Flat branch receives the required evidence.

\begin{wrapfigure}{r}{0.66\textwidth}
    \centering
    \includegraphics[width=\linewidth]{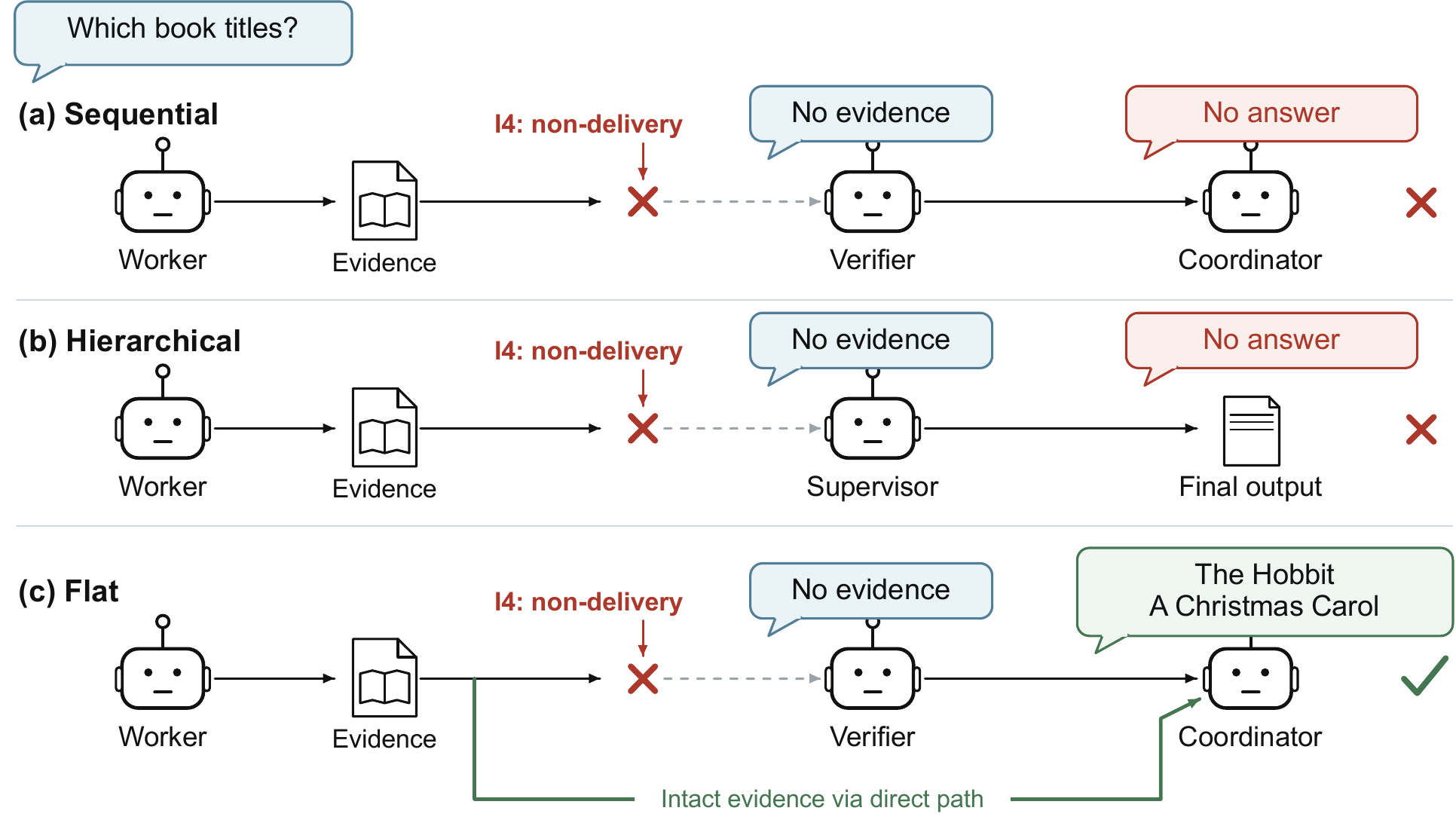}
    \caption{
    Representative topology-dependent outcome under the same I4 non-delivery fault.
Sequential and Hierarchical lose their only downstream evidence path, whereas
Flat preserves the required evidence through its additional direct branch.
    }
    \label{fig:rq3-topology-case}
\end{wrapfigure}
At the request stage, the success-rate difference remains within 0.5\%.
At the state-return stage, it remains within 0.7\% for non-delivery and 0.3\% for semantic corruption.
Sequential and Hierarchical also show similar behavior across these conditions.
The large Flat advantage therefore appears when its additional branch already contains the information required by downstream agents. 
The topology in \autoref{fig:topology-comparison} explains this pattern.
The Flat branch diverges after the Worker has generated the required evidence.
A fault at the evidence handoff can affect one route while the second route continues to carry the same evidence to the final decision.
Faults at the request and state return stages occur before this divergence.
Both Flat branches then depend on the same affected information.
\autoref{fig:rq3-topology-case} provides a trace-level example of this paired advantage.
Under the same I4 non-delivery fault, only Flat preserves the required evidence
through its direct branch.
This is consistent with dependability principles in which redundancy helps when
an alternative path avoids the affected dependency~\cite{avizienis2004dependability,alvaro2015ldfi}. The topology effect also remains stable across task compositions.
Removing each task context in turn leaves the net Sequential-to-Flat success-rate improvement between 17.6\% and 29.5\%.
The Hierarchical-to-Flat improvement remains between 17.5\% and 29.4\%.
The same direction holds when we use all executions with known outcomes and when we restrict the analysis to executions with successful matched clean runs.
Flat substantially reduces overall failures, while the remaining failures become increasingly concentrated in silent semantic faults.
In the detection-complete matched cohort, Sequential and Hierarchical fail on 94.6\% and 94.7\% of executions, respectively, compared with 32.1\% under Flat.
The silent-failure rate per execution changes only slightly, from 17.6\% under Sequential to 16.6\% under Flat.
Consequently, silent failures account for 51.6\% of the failures that remain under Flat, compared with 18.6\% under Sequential and 18.1\% under Hierarchical.
Semantic corruption and partial-content faults contribute 94.8\% of these residual silent failures.
The additional information path therefore removes many execution failures, while semantic faults remain the main residual risk.

\finding{\textbf{Flat improves fault tolerance when its additional path still carries the required evidence.}
Its success-rate gain reaches 66.5\% to 66.6\% for evidence-handoff non-delivery and 41.3\% to 42.3\% for evidence-handoff corruption.
The gain remains below 0.7\% when the affected communication occurs before the additional path receives the required evidence.}

\lead{Impact of model configuration.}
Model configuration changes the magnitude of fault impact, especially under Flat.
As shown in \autoref{fig:rq3-topology}(a), clean-conditional loss under Flat ranges from 36.8\% to 61.6\% across the three models.
Sequential and Hierarchical remain between 96.5\% and 97.2\% under the same model configurations.
Model choice therefore affects how well agents handle the information that remains available, while substantial communication-induced degradation persists across all three models. We further quantify the recovery provided by model changes using 1,408 common faulted units.
Changing the deployed model configuration recovers only 0.2\% to 4.5\% of baseline failures.
In comparison, changing the topology to Flat recovers 19.3\% to 27.0\%.
Across the 12 matched comparisons, the topology change provides an additional net success-rate gain of 17.5\% to 23.4\%. The failures recovered by the two changes also show little overlap, with Jaccard similarity ranging from 0 to 0.047.
This result indicates that model changes can recover a small set of executions that respond differently to model configuration.
Their overall recovery coverage, however, remains limited compared with the impact of changing the information path.

\finding{\textbf{Changing the LLM provides limited recovery from communication faults.}
Across the common mixed-fault cohort, changing the LLM recovers only 0.2\% to 4.5\% of baseline failures, while changing to Flat recovers 19.3\% to 27.0\%.
The recovered sets show little overlap, indicating that model changes affect a small and largely different subset of failures.}

\subsection{RQ4. Protection Mechanisms}
\label{sec:rq4}

RQ4 investigates how communication protection should respond to execution faults and information faults, and when recovery remains effective.
Following recent studies on fault tolerance and verification in LLM-based multi-agent systems~\cite{jia2026mas,cemri2025mast}, we organize the evaluated mechanisms into execution-side protection and application-level evidence protection.
Execution-side protection uses resubmission and duplicate suppression to preserve intended actions, while evidence protection checks task-relevant evidence and reacquires information when the received content is incomplete or inconsistent.
We then evaluate their effectiveness, recovery conditions, and execution overhead.

\begin{wrapfigure}{r}{0.5\textwidth}
    \centering
    \includegraphics[width=\linewidth]{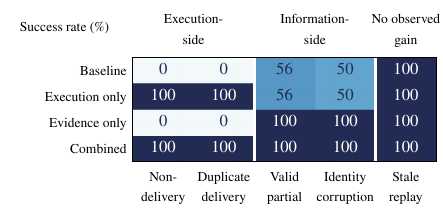}
    \caption{Success rates under the main protection study.
    Execution-side protection primarily recovers execution faults,
    whereas evidence protection recovers information faults.}
    \label{fig:mitigation-heatmap}
\end{wrapfigure}

\lead{Fault-specific protection and evidence recovery.}
The main protection study shows that execution faults and information faults benefit from different mechanisms.
Among 106 baseline failures, execution-side protection uniquely recovers 67.9\%, while evidence protection uniquely recovers the remaining 32.1\%.
The two recovery sets do not overlap.
\begin{figure}[t]
    \centering
    \includegraphics[width=0.9\columnwidth]{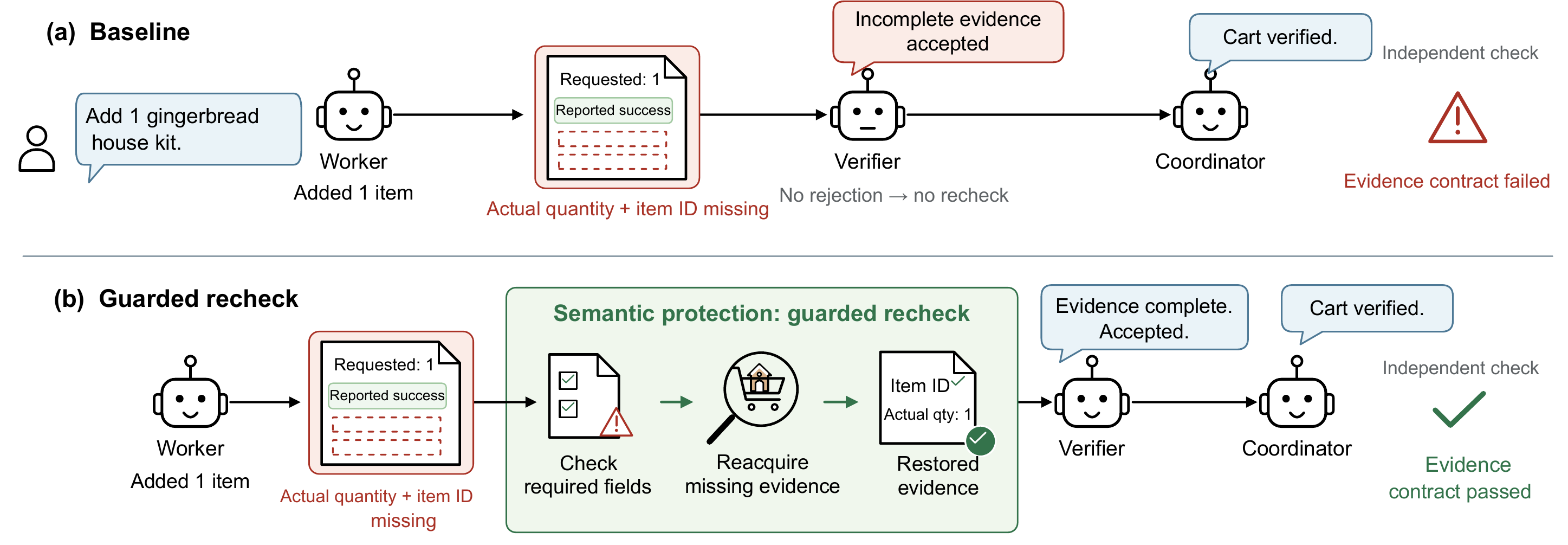}
    \caption{
    Guarded evidence recovery repairs incomplete evidence without repeating the completed action.
    }
    \label{fig:rq4-guarded-recheck}
\end{figure}
As shown in \autoref{fig:mitigation-heatmap}, execution-side protection raises success from 0\% to 100\% under the evaluated non-delivery and duplicate-delivery faults.
Its effect is small for the tested evidence faults.
Evidence protection shows the complementary pattern.
For valid partial content, success increases from 55.6\% to 100\%.
For identity corruption, success increases from 50.0\% to 100\%.
Combined protection covers both fault classes and reaches 100\% success across the five tested conditions.
Stale replay already achieves full task success in this task family, leaving no additional recovery gain for the protection mechanisms.
The targeted evidence-protection study explains how this recovery occurs.
Under Sequential, the baseline succeeds in only 5.6\% of the evaluated valid-partial and identity-corruption executions.
Evidence checking restores success when rejection leads to an available unaffected recovery path.
The check identifies evidence that violates the expected contract.
The recovery path then provides valid information for the downstream decision.
This separation distinguishes fault detection from information recovery.
Final task success can also hide errors inside the workflow.
Under Flat, the baseline completes all evaluated faulted executions successfully, while 97.2\% contain a recorded incorrect intermediate evidence acceptance.
Evidence protection preserves the final success rate and removes these recorded incorrect acceptances.
Protection therefore improves the internal evidence flow even when the final task outcome remains unchanged. 
\autoref{fig:rq4-guarded-recheck} illustrates this mechanism in a representative execution.
The same pattern appears in the targeted Reddit study.
Protection raises success from 44.4\% to 77.8\% for valid partial content and from 33.3\% to 77.8\% for semantic corruption, matching the contemporaneous clean success rate.
All protected faulted executions trigger fresh reacquisition, and 88.9\% subsequently adopt the new evidence.
The remaining failures persist after reacquisition, showing that recovery removes communication-induced degradation when trustworthy evidence can still be restored.

\lead{Recovery boundary and overhead.}
Recovery depends on whether the workflow can still obtain unaffected evidence.
The source-level analysis makes this requirement explicit.
Forwarding another copy of corrupted evidence preserves the same incorrect information, whereas independent reacquisition obtains a fresh observation for downstream agents.
Additional copies therefore help only when they preserve access to information outside the affected fault path.
The repeated-exposure study further identifies this recovery boundary.
As summarized in \autoref{tab:recovery-boundary}, Combined protection restores all Sequential failures when corruption is limited to the initial evidence handoff and the recovery delivery remains unaffected.
Recovery disappears when the same fault also reaches subsequent recovery deliveries.
Flat remains successful under this broader exposure because its additional evidence path remains available.
When the fault also reaches this path, Flat can no longer recover.
The environment action remains successful across these conditions, showing that the final failures arise from unavailable trustworthy evidence rather than failed environment execution.
\begin{wraptable}{r}{0.44\textwidth}
  \centering
  \caption{Recovery boundary under repeated fault exposure.
  Entries show Baseline $\rightarrow$ Combined successful runs.}
  \label{tab:recovery-boundary}
  \small
  \setlength{\tabcolsep}{3pt}
  \renewcommand{\arraystretch}{1.1}
  \begin{tabular}{@{}lcc@{}}
    \toprule
    Fault scope & Sequential & Flat \\
    \midrule
    Initial handoff
      & $0\to18$ & $18\to18$ \\
    \shortstack[l]{Handoff and\\recovery deliveries}
      & $0\to0$  & $18\to18$ \\
    \shortstack[l]{All evidence\\paths}
      & $0\to0$  & $0\to0$ \\
    \bottomrule
  \end{tabular}

  \smallskip
  \parbox{\linewidth}{\footnotesize
  Each condition and protection arm contains 18 runs.}
\end{wraptable}
Conditional reacquisition also reduces unnecessary communication during clean execution.
Guarded readback triggers no dedicated reacquisition in the evaluated clean runs, whereas always-on readback triggers it in every run.
Guarded readback removes two workflow HTTP requests per clean execution.
This corresponds to a 15.4\% reduction under Sequential and a 13.3\% reduction under Flat, with no change in task success or model-call count.
Under the evaluated fault conditions, both strategies invoke reacquisition when recovery is required.
Guarded recovery therefore concentrates the additional communication cost on executions that fail the evidence check.

\finding{\textbf{Effective protection must match the failed dependency and retain access to trustworthy evidence.}
Execution-side and evidence protection recover complementary failure sets.
Evidence recovery remains effective while an unaffected evidence path is available, but fails when the same fault reaches every available path.
Guarded reacquisition preserves this recovery capability while reducing unnecessary communication during clean execution.}

\section{Discussion}
\label{sec:discussion}

\subsection{Implications}
\label{sec:implications}

\noindent\textbf{\textcircled{\scriptsize 1} Information dependency.}
Our results show that the impact of a communication fault depends strongly on the affected information dependency. The same fault can produce different outcomes across communication stages and task contexts, depending on whether later execution can obtain equivalent information.
\noindent\textbf{\textcircled{\scriptsize 2} Topology and redundancy.}
Workflow topology improves fault tolerance when it preserves an unaffected information path. This explains why Flat provides large gains at evidence handoff but little benefit when the fault occurs before its additional branch receives the required information.
\noindent\textbf{\textcircled{\scriptsize 3} Silent information faults.}
Structural redundancy removes many overt execution failures, but it does not eliminate incomplete or corrupted information that remains processable. These faults can therefore account for a larger share of the remaining failures. Reliable debugging and protection should consider both message delivery and the trustworthiness of information consumed downstream.

\subsection{Future Work}
\label{sec:future-work}

\noindent\textbf{\textcircled{\scriptsize 1} Broader systems and environments.}
Future work can extend the study to additional agent frameworks, workflow
topologies, communication protocols, and controlled model configurations.
Such evaluations could clarify which observations generalize across systems
and which depend on specific runtimes.
\noindent\textbf{\textcircled{\scriptsize 2} Task-centric sensitivity.}
Characterizing tasks by evidence depth, recoverable state, alternative
information paths, and critical dependencies may better explain fault
sensitivity than dataset identity alone.
This could also support task-aware fault prediction and protection design.
\noindent\textbf{\textcircled{\scriptsize 3} Adaptive recovery.}
Finer-grained instrumentation of detection and recovery could support adaptive
protection that selects resubmission, alternate routing, evidence checking, or
fresh reacquisition according to the affected dependency.
Evaluating these mechanisms under persistent or correlated faults could further
identify the limits of redundancy and recovery.
Such analysis may also reveal when recovery itself becomes a new source of
communication overhead or failure.

\subsection{Threats to Validity}
\label{sec:threats}

\noindent\textbf{\textcircled{\scriptsize 1} Construct validity.}
Our measurements may not capture every stage of fault detection and recovery. We reduce this threat by separating receiver-visible exposure ($A$), downstream MAS consequence ($M$), and final task outcome ($O$), while also recording the handling-response flag $B$ and trace-linked state changes. We assign semantic consequences only when altered information is accepted or used downstream, and we do not interpret $B$ as evidence of complete detection, correction, or state restoration.
\noindent\textbf{\textcircled{\scriptsize 2} Internal validity.}
Fault injection may introduce artifacts if an intervention does not represent a plausible communication fault or affect the intended dependency. We ground fault mappings in communication standards and prior studies, validate reproducible lower-stack mappings against their expected receiver-visible manifestations, verify each applied intervention in execution traces, and use matched clean and faulted executions to isolate fault-induced effects. Some physical and hardware-dependent mechanisms require conditions beyond our testbed, so we do not claim direct experimental reproduction of every underlying mechanism.
\noindent\textbf{\textcircled{\scriptsize 3} External validity.}
We evaluate 100 tasks from five domains, three AutoGen topologies, and three deployed model configurations. The observed effects may still depend on the evaluated communication paths, task contracts, runtime mechanisms, and framework implementation, so we restrict our conclusions to the evaluated settings.
\noindent\textbf{\textcircled{\scriptsize 4} Conclusion validity.}
LLM-MAS executions contain stochastic variation that may affect observed outcomes. We reduce this threat through repeated executions, matched comparisons, task-clustered statistical analyses, and uncertainty estimates reported with the quantitative results.

\section{Conclusion}

This work presents a cross-layer taxonomy and fault-injection pipeline for studying communication faults in LLM-based multi-agent systems. Across 15,882 applied faults, 89.4\% propagate to an MAS consequence and 80.2\% end in task failure, while some propagated faults still complete successfully, showing that manifestation, propagation, and final outcome capture different aspects of reliability. Our results further show that fault impact depends on the affected dependency, task context, workflow topology, and deployed model configuration. Protection must therefore match the failed dependency: execution faults benefit from action-side handling and path redundancy, whereas corrupted or incomplete information requires semantic checking and, in some cases, independent reacquisition. Reliable LLM-MAS communication requires protecting both message delivery and the trustworthiness of the information consumed downstream.

\section{Data Availability}
\label{sec:data}

Our artifact provides source code, processed results, analysis configurations,
operator mappings, trace/adjudication definitions, and scripts for reproducing
the reported figures and tables. The artifact repository is available at
\href{https://github.com/SonGoKu21/mas_communication/}{MAS-Communication},
with a companion website at
\href{https://songoku21.github.io/mas_communication/}{MAS-\allowbreak Communication-\allowbreak Website}.
Raw execution logs, complete task environments, credentials, private content,
and private model reasoning are not included.

\bibliographystyle{ACM-Reference-Format}
\bibliography{references}

\end{document}